\documentclass[12pt,preprint]{aastex}

\def\ifundefined#1{\expandafter\ifx\csname#1\endcsname\relax}

\newif\ifpdf
\ifx\pdfoutput\undefined
   \pdffalse      
\else
   \pdfoutput=1   
   \pdftrue
\fi

\def\la{\mathrel{\hbox{\rlap{\hbox{\lower4pt\hbox{$\sim$}}}\hbox{$<$}}}}
\def\ga{\mathrel{\hbox{\rlap{\hbox{\lower4pt\hbox{$\sim$}}}\hbox{$>$}}}}

\newcommand{\be}{\begin{eqnarray}}
\newcommand{\ee}{\end{eqnarray}}

\ifundefined{ensuremath}\def\ensuremath#1{\relax\ifmmode{#1}}
\else${#1}$\fi\else\relax\fi
\ifundefined{nuc}\def\nuc#1#2{\relax\ifmmode{}^{#1}{\protect\text{#2}}
\else${}^{#1}$#2\fi}\else\relax\fi

\newcommand{\etal}{et al.}

\newcommand{\kmps}{km~s$^{-1}$}

\newcommand{\msol}{\ensuremath{{\textrm{M}_\odot}}}

\newcommand{\nni}{\nuc{56}{Ni}}

\def\ang{\hbox{\AA}}

\def\tstd{\ensuremath{\tau_{\textrm{std}}}}

\newcommand{\phx}{\texttt{PHOENIX}}

\newcommand{\sneia}{SNe~I\lowercase{a}}
\newcommand{\gamray}{$\gamma$-ray}

\received{} 
\accepted{} 
\journalid{}{} 
\articleid{}{} 
\shortauthors{Baron, E. et~al.}
\shorttitle{Detectibility of C+O in SNe Ia}

\begin{document}

\title{Detectibility of Mixed Unburnt C+O in Type I\lowercase{a} Supernova
Spectra}

\author{ E.~Baron\altaffilmark{1}\email{baron@nhn.ou.edu}  Eric
J.~Lentz\altaffilmark{2}\email{lentz@physast.uga.edu}  Peter
H.~Hauschildt\altaffilmark{3}\email{phauschildt@hs.uni-hamburg.de}
}

\altaffiltext{1}{Department of Physics and Astronomy, University of
Oklahoma, 440 West Brooks, Rm.~131, Norman, OK 73019, USA}

\altaffiltext{2}{Department of Physics and Astronomy \& Center for
Simulational Physics, University of Georgia, Athens, GA 30602, USA}

\altaffiltext{3}{Hamburger Sternwarte, Gojenbergsweg 112,
21029 Hamburg, Germany}

\begin{abstract}
Motivated by recent 3-D calculations of the explosion of Type Ia
supernova via a pure deflagration we calculate the observed spectra at
15--25 days past maximum light of a parameterized model which has a
considerable fraction of unburnt C+O in the central regions. Rather
than attempting a self consistent 3-D calculation, which is beyond the
scope of current computer codes, we modify the composition structure
of the 1-D deflagration model W7. In our exploratory
parameterized calculations, we find that a central concentration of C+O is not
ruled out by observations for the epochs we study. We briefly examine
whether nebular phase spectra could be incompatible with observations.
\end{abstract}

\section{Introduction}
Interest in Type Ia supernovae has been high over the last twenty
years, since it was recognized that they closely meet the
astronomer's criteria for a ``standard candle''. This interest has
grown with the realization that one can use the observed light curve
to make a luminosity correction \citep[similar to the Cepheid P-L
relation,][]{philm15,rpk95,rpk96,philetal99}. SNe~Ia played an
important role in current determinations of the Hubble constant with
the \emph{Hubble Space Telescope} \citep{sandage90N96,brancharaa98}
With the discovery of the ``dark energy''
\citep{riess_scoop98,perletal99,goldhetal01} enthusiasm for studies of
Type Ia supernovae has grown. Heretofore, all calibrations of the
luminosity of SNe~Ia have been empirical in nature; relying largely on
the sample obtained by the Calan-Tololo survey \citep{hametal96b}. While
significant efforts are underway to improve the local sample of
SNe~Ia, theoretical  understanding of the explosion mechanism and
progenitor system are required in order to produce confidence in the
validity of the empirically based results.

In spite of the fact that the progenitor system for SNe~Ia is still
uncertain \citep{prog95}, there is little doubt that it requires the
thermonuclear explosion of a C+O
white dwarf. Given this fact, progress is being made in the understanding
of the explosion mechanism itself and calculations have progressed
from 1-D and 2-D, to full 3-D
\citep{khok93,khokh01,hnar00,RHN99,RNH02,RHN02a,RHN02b,Gamezoetal03} 

The calculation of the explosion mechanism is a formidable numerical
problem. There is good agreement that the explosion begins
as a deflagration with a laminar flame, which then develops into a
fully turbulent flame via various instabilities, dominated by
Rayleigh-Taylor instabilities.  Even in 3-D,
the flame cannot be fully resolved and a variety of subgrid models
have been developed. Recent work has focused on calculations of the
deflagration of C+O Chandrasekhar mass white dwarfs.
While some calculations have had trouble obtaining the observed
energy of the explosion, there seems to be general agreement that a
fully resolved 3-D calculation can produce significant kinetic energy
and enough nickel  to reproduce the general features of a Type Ia
explosion. Recently, \citet{Gamezoetal03} performed a series of 3-D
deflagration model calculations in which they obtained a total kinetic
energy of about $10^{51}$~ergs. The key feature of the simulations is
the highly convoluted turbulent flame surface that allows extensive
interpenetration of burnt and unburnt materials. Similar results were
found by \citet{RHN02a,RHN02b}. However, the strong interpenetration
of burnt and unburnt material leaves a significant amount of unburnt
C+O in the central regions of the ejecta.  This result led
\citet{Gamezoetal03} to conclude that deflagration models would be in
conflict with 
 observed supernova spectra because there would be 
strong carbon or oxygen lines produced by the model,
which are not seen in observed SNe~Ia spectra. Based on this inference,
\citet{Gamezoetal03} rule out the deflagration model completely as a
viable model for the explosion mechanism of SNe~Ia. In drawing this
conclusion \citet{Gamezoetal03} rely on the work of
\citet{jeffetal92}, \citet{fish90n97}, \citet{mazz90N01}, and
\citet{kir92a} which all 
use a highly parameterized direct spectral analysis technique for
producing synthetic spectra. In particular, all these analyses make
the Schuster-Schwarzschild approximation and thus cannot address the
formation of lines below the imposed photosphere.  These analyses are
extremely useful for 
line identifications in observed spectra and for determining the
velocity interval at which certain ions \emph{must} exist. However,
these analyses cannot (and do not claim to) determine whether a given
species must be absent. That is, from these analyses one can conclude
that a certain ion must be present in a certain velocity interval, but
one cannot conclude that any particular ion must be absent. 

In order to examine the true viability of a full 3-D hydrodynamical
calculation, a full 3-D radiative transfer calculation should be
performed. However sophisticated 3-D radiative transfer codes do not
yet exist that can tackle the problem of detailed radiative transfer
in SNe~Ia. In order to examine the viability of deflagration models we
present the results from \phx\ calculations using a parameterized
version of the tuned 1-D deflagration model W7.

\section{Calculations}

The W7 model \citep{nomw7,nomw72} is a good match to normal \sneia\
spectra \citep{harkness91a,harkness91b,nug1a95,nugseq95,whk95,l94d01},
especially the outer layers visible during the 
pre-maximum phase of the supernova. The outer layers of W7 are
unburnt C+O enriched with solar abundance metals, expanding at
velocities, $v > 
15000$ \kmps, with a total mass of $\sim 0.07$~\msol, followed by
intermediate mass elements, with an interior composed mostly of \nni\
but also including some other iron group elements.  The model
is a 1-D deflagration model, where the flame speed was adjusted to
produce intermediate mass elements at high velocity as observed in
SNe~Ia spectra.

We have altered the inner composition of the W7 to produce two 
alternate models with  C+O mixed into the center.  The models are 
prepared by removing a portion of the original mass in the W7 model and 
replacing it with an equal mass of C+O, with the C/O ratio = 1 by 
number, keeping the density structure the same. The C+O is used for the 
unburnt fuel mixed to the core found in 3-D deflagration models.  In 
the first model, labeled ``25\% C+O core'', we have replaced 25\% of the 
inner core with C+O. The mixed material extends from the center to 
3000~\kmps, where the mixed fraction is then reduced linearly in 
velocity space to 0\% at 4000~\kmps.  In the second model, designed to 
mock-up the \citet{Gamezoetal03} model and labeled ``50\% C+O core'',
50\% of the mass  
is replaced with C+O from the center to 4000~\kmps.  The mixed fraction 
is reduced linearly in velocity space to 10\% at 5000~\kmps.  The 10\% 
C+O region extends to 15000~\kmps\ where the W7 model becomes entirely 
unburnt material.
Examining Figure~3 of
\citet{Gamezoetal03} it is diffiult to determine the exact
composition structure, since they only distinguish between unburnt
(C+O) and ``burnt'' material, but the exact composition of the burnt
material is unspecified because it is too computationally intensive
to follow the reaction rates in detail.

The spectra and \gamray\ deposition are calculated using the general
purpose radiative transfer code \phx, version {\tt 11.9}
\citep{hbjcam99}. There is some change in the total gamma-ray
deposition with the composition change, but the change is smaller than
one might expect, since in this velocity range most of the material in
W7 is neutron-rich iron peak elements and not \nni. For the 25\% C+O
core model the change is 2\%, and for the 50\% C+O core model the
change is 20\% at both epochs. \phx\ includes all of the effects of
special relativity in the steady-state solution of the transport
equation and energy balance.  For these exploratory calculations we
have treated all species in LTE. Studies by \citet{snefe296} and
\citet{whhs98} have shown the LTE is a reasonable approximation in
SNe~Ia and since we are looking only qualitatively for lines that are
clearly absent in observed spectra the assumption of LTE should not
affect our conclusions. The use of \phx\ with W7 models is described
in detail in \citet{l94d01}.

\section{Results}

\begin{figure}
\begin{center}
\includegraphics[width=12cm,angle=90]{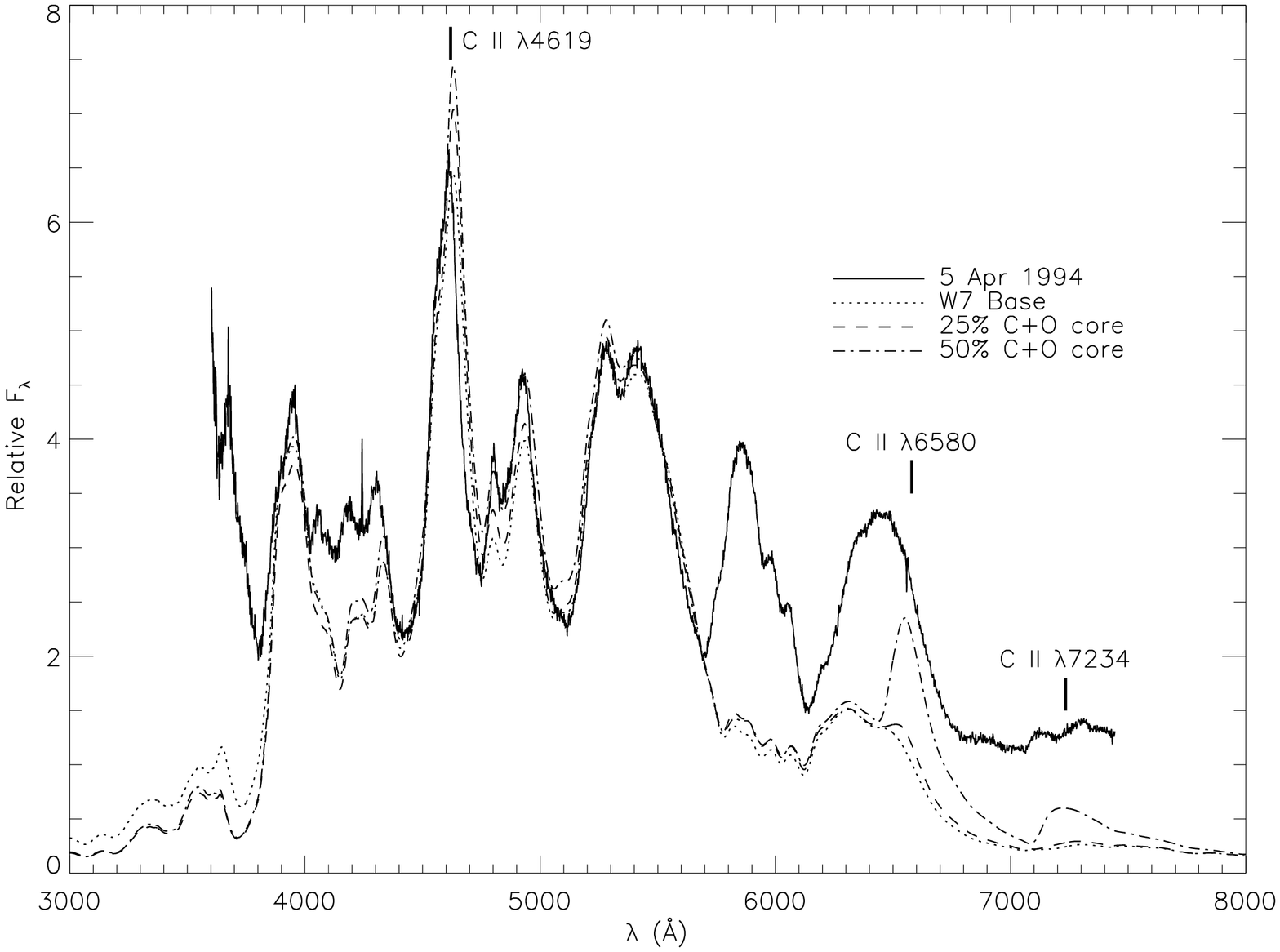}
\end{center}
\caption{\label{fig:d35_composite} The observed spectrum of SN~1994D
obtained on 1994 April 5\citep{patat94D96} is compared to synthetic
spectra computed assuming this corresponds to 35~days past explosion
using W7 compositions (base), W7 with C+O mixing only into the
central regions (25\% C+O core) and C+O in the core and intermediate velocity
regions (50\% C+O core). The line identifications are to guide the
eye, the actual features are blends.}
\end{figure}

Figure~\ref{fig:d35_composite} displays the observed spectrum of
SN~1994D on  1994 April 14, roughly 15 days past maximum light \citep[35
days past explosion,][]{patat94D96}  and the results of three calculations:
1) the original W7 model (denoted base), 2) the ``25\% C+O  core'' model,
and 3) the ``50\% C+O core'' model.

\begin{figure}
\begin{center}
\includegraphics[width=12cm,angle=0]{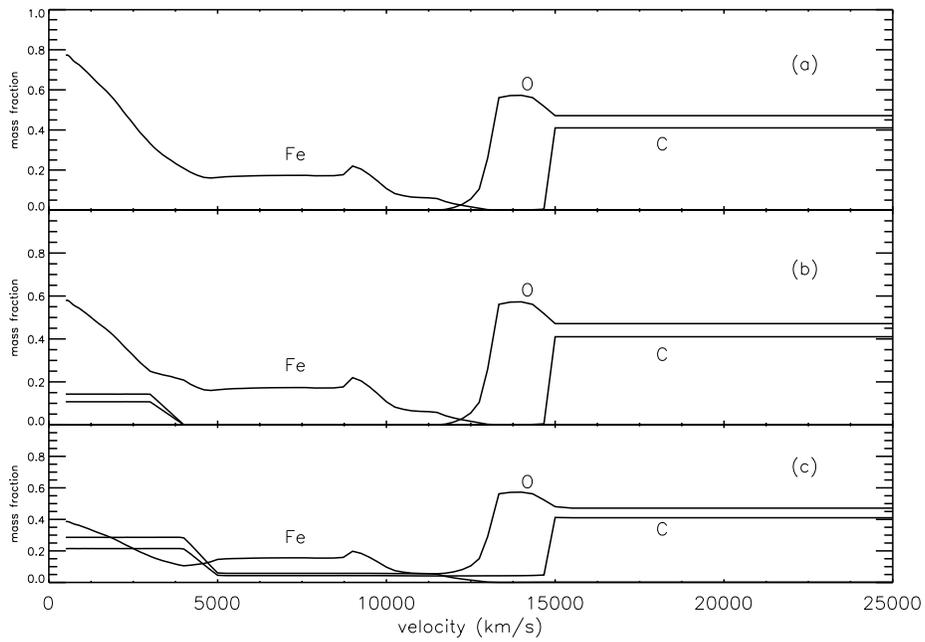}
\end{center}
\caption{\label{fig:compare_composition} The unmodified (base) W7
compositions as a function of velocity are shown in panel (a),  the
core only  mixed compositions (25\% C+O core) are shown in panel (b), and the
``50\% C+O core'' compositions are shown in panel (c).}
\end{figure}

Figure~\ref{fig:compare_composition} displays the 
composition structure of the three models at day 35.  While
\ion{C}{2} lines become more prominent they do not necessarily rule
out the existence of a substantial amount of C+O. 
Fig.~\ref{fig:d35_composite} 
shows that as the C+O abundance is increased, the \ion{C}{2} $\lambda
4619$ line becomes stronger, as does the \ion{C}{2} $\lambda6580$
line, and the \ion{C}{2} $\lambda7234$ line. 
In the UV there is no evidence for, e.g., \ion{C}{4}. 
In the ``25 \% C+O core'' model, C+O mixing has not reached its peak until
$\tstd = 0.76$ 
(the total continuum optical depth at 5000~\ang). 

\begin{figure}
\begin{center}
\includegraphics[width=12cm,angle=90]{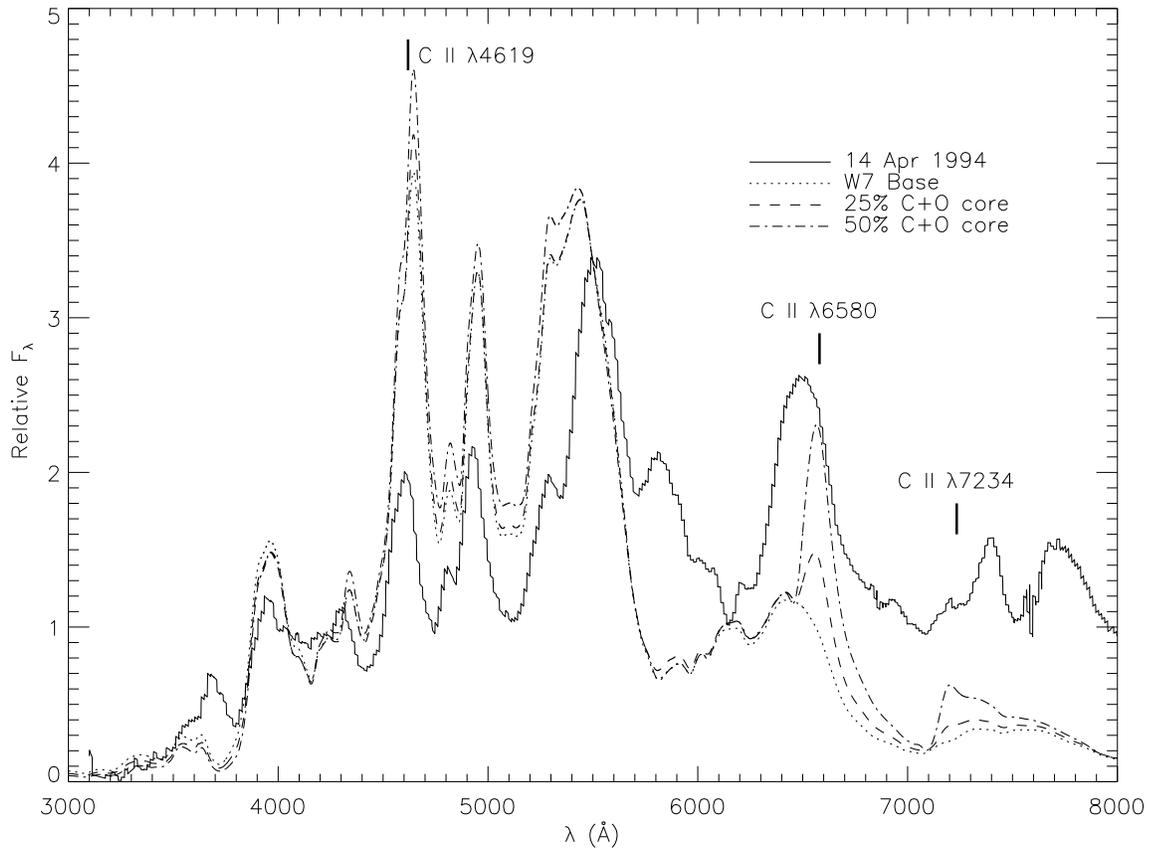}
\end{center}
\caption{\label{fig:d44_composite} The observed spectrum of SN~1994D
obtained on  1994 April 14 \citep{patat94D96} is compared to synthetic
spectra computed assuming this corresponds to 44~days past explosion
using W7 compositions (baseline), 25\% C+O core, and 50\% C+O core
compositions.} 
\end{figure}

To verify that we
are indeed sampling the line forming region with our mixed models we
calculated a model 44 days after explosion (roughly 24 days after
maximum light) which is shown in Figure~\ref{fig:d44_composite}. In
this case, $\tstd =1$ at 1100, 1450, 2000~\kmps\ in the base model, the
25\% C+O model, and the 50\% C+O core model respectively, since O
serves as a source of free electrons. At this
epoch the same features 
become too strong, although since our models are pure LTE and they use
only the \citet{kurucz93} line list which does not include forbidden
lines. Therefore, lines  of [\ion{Fe}{3}] and
[\ion{Co}{3}]  which are generally assumed to produce the features near
5500~\ang\ and 6500~\ang\ \citep{bowersetal97} are not included in our
calculations. It is interesting to note that the velocity of the
absorption minimum of the \ion{C}{2} $\lambda6580$ line in our model
spectra is at about $6000$~\kmps, which is where \ion{C}{2} is both
prevalent, in our 50\% C+O core model, and is the top of the canonical
line forming 
region $\tstd = 0.1$. In the 25\% C+O core model, the minimum moves
somewhat redward to about $5400$~\kmps. However in this model there is
no carbon between 4000 and 15000 \kmps\ so the minimum is likely
produced by a complicated blend of lines as is almost always the
case in detailed models. The peak is almost certainly produced by
\ion{C}{2} $\lambda6580$. 

The models of \citet{Gamezoetal03}
have both C+O mixing to low velocity, and also mixing of burnt
material (possibly iron) to high
velocity and we have not attempted to simulate that here. In addition
the results of \citet{Gamezoetal03} end at 1.9~s, before the explosion
has become homologous, so it is difficult to translate their plot of
compositions as a function of radial coordinate into a model in
homologous expansion where both velocity and radius are equally good
radial coordinates.

\section{Discussion} 

Our results clearly demonstrate that one cannot simply examine the
radial composition structure of a model and determine whether a
hydrodynamical model is viable or not. On the other hand, we do claim
to have shown that the model of \citet{Gamezoetal03} \emph{is} a
viable model for SNe~Ia. In particular, it is not possible to
determine the exact composition structure from their paper; future
work should attempt to make some educated guesses about the
composition structure and calculate synthetic spectra directly from
their (angle averaged) model. \citet{SB03a} calculated multi-band
light curves from the model of \citet{RHN02b} and found that having
burnt products at high velocity was somewhat helpful particularly in
$U$ and $B$. This is an indication that mixed models may be viable for
SNe~Ia, whether produced by deflagration or some other
mechanism. Nevertheless, a full synthetic spectrum calculation is
required before firm conclusions can be drawn about the viability or
lack thereof of any particular hydrodynamical model. What we have
shown is that unburnt C+O in the central regions are not necessarily
ruled out at the epochs we study.

One possible problem with the model of \citet{Gamezoetal03} is that
late time nebular spectra would show strong forbidden oxygen lines,
which have not typically been observed in SNe~Ia, although
\citet{mink37C39} did identify weak narrow lines in SN~1937C of
[\ion{O}{1}] $\lambda 6300$ and $\lambda 6394$ beginning 184~d after
maximum. The lines were absent by 275~d after maximum. Again, until
detailed late-time synthetic spectra have been calculated, it is
impossible to draw a firm conclusion. It may be that since the
material in the center of the \citet{Gamezoetal03} model is dense and
the radioactive \nni\ is well mixed with the C+O that the oxygen stays
ionized for a long time, but it does seem likely that [\ion{O}{1}]
lines should appear at some point.

In addition, the models of \citet{Gamezoetal03} mix burnt material
into the outer layers as well as mixing unburnt material into the
inner layers. We have not accounted for that effect in these simple
exploratory calculations. However, we know that SNe~Ia show a range of
``temperatures'', forming a ``temperature sequence'' that correlates
with the light curve shape sequence \citep[][S.~Bongard \etal, in
preparation]{nugseq95}. Exactly how the sequence is produced is not
understood, although it is clear that models that produce more nickel
will be both hotter and have higher opacity than those with less
nickel \citep{kmh93}. ``Hot'' SNe~Ia show lines of \ion{Fe}{3}
\citep{fisher91T99,hat91T02} and \ion{Si}{3} (D.~Branch \etal, in
preparation), but in parameterized models these lines are produced via
extra thermalization \citep{nugseq95} and not by direct excitation due
to non-thermal electrons produced by \nni\ decay. If very
inhomogeneous mixing of burnt and unburnt material in the outer
layers is a general feature of deflagration models \citep{khokh01},
then the observed uniformity of the \ion{Si}{2} 6150\ang\ feature
probably rules these models out \citep{thomas02}.

From our simple parameterized results, we conclude that the viability
or non-viability of any hydrodynamical model  cannot simply be
determined by the presence of a particular element in a certain
velocity interval. The converse is not true, the absence of a
particular element in a velocity interval where it is observed is
fatal to any particular hydrodynamical model. It will be helpful when
future hydrodynamical simulations can extend to the regime of
homologous expansion and when the full compositions structure of the
models are available.

\acknowledgments We thank David Branch for very useful discussions and
significantly improving the presentation of this work. This
work was supported in part by NASA grant 
NAG5-12127, NSF grant AST-0204771, and an IBM SUR grant to the
University of Oklahoma; and by NSF grant AST-9720704, NASA ATP grant
NAG 5-8425 and LTSA grant NAG 5-3619 to the University of Georgia.
PHH was supported in part by the P\^ole Scientifique de Mod\'elisation
Num\'erique at ENS-Lyon. Some of the calculations presented in this
paper were performed at the San Diego Supercomputer Center (SDSC),
supported by the NSF, on the IBM pSeries 690 of the Norddeutscher
Verbund f\"ur Hoch- und H\"ochstleistungsrechnen (HLRN), and at the
National Energy Research Supercomputer Center (NERSC), supported by
the U.S. DOE. We thank all of these institutions for a generous
allocation of computer time.


\clearpage

\end{document}